\documentclass[aps,a4paper,twocolumn,showpacs,showkeys,preprintnumbers,amsmath,nofootinbib,amssymb,10pt]{revtex4}

\usepackage{graphicx}
\usepackage{amsmath}
\usepackage{amsthm}
\usepackage{amsfonts}
\usepackage{amssymb}

\usepackage{xcolor}

\colorlet{linkequation}{blue}
\usepackage[colorlinks]{hyperref}

\newcommand{\be}{\begin{equation}}
\newcommand{\ee}{\end{equation}}
\newcommand{\bea}{\begin{eqnarray}}
\newcommand{\eea}{\end{eqnarray}}
\newcommand{\beas}{\begin{eqnarray*}}
\newcommand{\eeas}{\end{eqnarray*}}

\newcommand{\mean}[1]{\left \langle #1 \right\rangle}
\usepackage[top=2.1cm, bottom=2.4cm,left=1.5cm, right=1.5cm,headheight=1.5cm,headsep=0.8cm]{geometry}

\begin{document}

\title{Optimal growth trajectories with finite carrying capacity}
\author{F. Caravelli$\ ^{1,2,3}$, L. Sindoni$\ ^{4}$, F. Caccioli$\ ^{3,5}$, C. Ududec$\ ^1$}
\affiliation{
$\ ^1$ Invenia Labs, 27 Parkside Place, Cambridge, CB1 1HQ, UK\\
$\ ^2$ London Institute of Mathematical Sciences, 35a South Street, London W1K 2XF, UK\\
$\ ^3$ Department of Computer Science, University College London, Gower Street, London WC1E 6BT, UK \\
$\ ^4$ Max Planck Institute for Gravitational Physics, Albert Einstein Institute,
Am M\"{u}hlenberg 1, 14467 Golm, Germany\\
$\ ^5$ Systemic Risk Centre, London School of Economics and Political Sciences, London, UK\\
}
\begin{abstract}
We consider the problem of finding optimal strategies that maximize the average growth-rate of  multiplicative stochastic processes. For a geometric Brownian motion the problem is solved through the so-called Kelly criterion, according to which the optimal growth rate is achieved by investing a constant  given fraction of resources at any step of the dynamics. We generalize these finding to the case of dynamical equations with finite carrying capacity, which can find applications in biology, mathematical ecology, and finance.
We formulate the problem in terms of a stochastic process with multiplicative noise and a non-linear drift term that is determined by the specific functional form of carrying capacity. We solve the stochastic equation for two classes of carrying capacity functions (power laws and logarithmic), and in both cases compute optimal trajectories of the control parameter. We further test the validity of our analytical results using numerical simulations. \ \\\ \\

\end{abstract}
\maketitle

%\section{}
%\subsection{}
\section{Introduction}
There are many interesting connections between statistical mechanics and the theory of financial markets \cite{Geanakoplos1997,BouchaudPotters2000}.
For instance, it has been stressed recently that the lack of ergodicity in the geometric Brownian motion process \cite{Peters1} has important implications for the optimal leverage problem, i.e. the problem of finding how much of a portfolio should be re-invested over time to maximize the logarithmic growth-rate of capital \cite{Peters2}.

For multiplicative processes, such as the geometric Brownian motion,
the effective growth-rate is not given by the drift term alone. More precisely, consider the process described by
\begin{equation}
dK(t)=\rho K(t)(\mu dt+\sigma dW(t)),
\end{equation}
with $\mu$ the drift of the stochastic process, $\sigma$ the noise amplitude, and $\rho$ a positive constant. 
Here $K(t)$ represents the capital of an investor at time t, while $\rho$ is the fraction of capital that is invested in a risky security, also known as the leverage. 
Using Ito's formula it can be shown that $\mean{d K(t)/K(t)}=\mu\rho dt$, while $\mean{d \log(K(t))}=\rho (\mu-\frac{\sigma^2}{2})dt$, where $\langle\cdot\rangle$ represents the ensemble average over the stochastic process $dW(t)$. The fact that these two expected values are different has been interpreted in \cite{Peters2,Peters3,LauLubensky2007} as a characteristic signature of the absence of ergodicity, since the first expression can be identified as an ensemble average, while the second can be seen as the time average over an infinitely long single instance of the stochastic process.

In this context it has been argued that maximizing expected log-return of \$ $K(t)$, often called the Kelly criterion \cite{Kelly1956,Kestner2003,Thorp2006,Breiman1961, opttrad1, opttrad2, CoverThomas1991,HanochLevy1969}, is a better objective in the long run than simply maximizing its average. For the case of the geometric Brownian motion, in \cite{Peters2} Peters discusses the differences between these two objectives, and provides a new interpretation of the optimal leverage obtained from the Kelly criterion by Thorp \cite{Thorp2006}, and shows that it is given by
\begin{equation}
\rho_{Thorp}=\frac{\mu}{\sigma^2}.
\end{equation}

In this paper, we extend this analysis by computing optimal growth trajectories in the case of  a multiplicative random process with finite carrying capacity, when the drift term $\hat\mu\equiv\mu(\rho K)$ is a decreasing function of  $\rho(t)K(t)$. 
A finite carrying capacity can be associated with the presence of market frictions, such as transaction costs. Although we frame it in terms of investment decisions, our analysis is of interest beyond finance. For instance, stochastic processes with finite carrying capacity are commonly used in biology to describe the growth of a population constrained by a finite amount of resources and in a random environment. Stochastic Gompertzian differential equations are for instance well known in population ecology \cite{Kot}, where one might want to control the amount 
or resources in order to optimize the growth of a population. This problem can be attacked using the methodology developed in this paper.

Below, we compute the optimal parameter $\rho(t)$ for logarithmic and power law functional forms of $\hat\mu\equiv\mu(\rho K)$. We will exactly solve the two models, and evaluate the optimal strategy, i.e. the parameter $\rho(t)$. In addition, we provide a methodology for evaluating the optimal control parameter $\rho(t)$ for generic series expansions of the carrying capacity parameter.  Finally we show that numerical simulations agree with our analytical results.

\section{The model}
%\subsection{Return functional}

Consider the following process: At time $t=0$, an investor has $K(0)= \$ K_0$, and, at any discrete time step $t$, must decide the fraction $\rho(t)$ of her capital to invest in a risky asset. At the end of each period, the risky asset pays a return $r(t)$, which is drawn from a Gaussian distribution with average $\mu$ and standard deviation $\sigma$. Here we assume that there is a transaction cost $c(t)$ per dollar associated with the purchase of  the risky asset, and that the asset purchased at time $t$ cannot be carried over to the next period, but needs to be sold at the end of each period. This is inspired by possible applications to wholesale electricity markets \cite{Wolak1}, in which trades have to be closed at the end of each trading day.
Furthermore, we assume that the remaining fraction, $1-\rho(t)$, of the capital is not invested, and that its value does not change over the day (i.e., we assume the risk free interest rate is $0$ for simplicity).

Under the specifications above, the capital evolves between time $t$ and $t+1$ as:
\begin{equation}
K(t+1) = \left[(r(t)-c(t)) \rho(t) +1 \right]K(t).
\end{equation}

If we are interested in the evolution of wealth over time horizons that are much longer than a trading period, we can consider the process in the limit of continuous time:
\begin{equation}
dK(t) = \rho(t)  K(t) (r(t)-c(t)).
\end{equation}
If we now assume that the returns evolve according to the stochastic process
\be
r(t) = \mu dt +\sigma dW_t,
\ee
and we define $c(t) = f\left(\frac{\rho(t) K(t)}{\tilde K}\right) dt$,
we can write 
\bea
\frac{dK(t)}{ K(t)} = \rho(t) \hat\mu(\rho(t)K(t),t) dt +\rho(t) K(t)\sigma dW_t,
\label{eq:capitalEvolution}
\eea
where 
\be
\hat\mu(\rho(t)K(t),t) =\mu \left[1-f\left(\frac{\rho(t) K(t)}{\tilde K}\right)\right] .
\ee
Here $\tilde K$ is the carrying capacity of the system, which in our context is associated with the cost of purchasing risky assets.

\begin{figure}
\includegraphics[width=8.5cm]{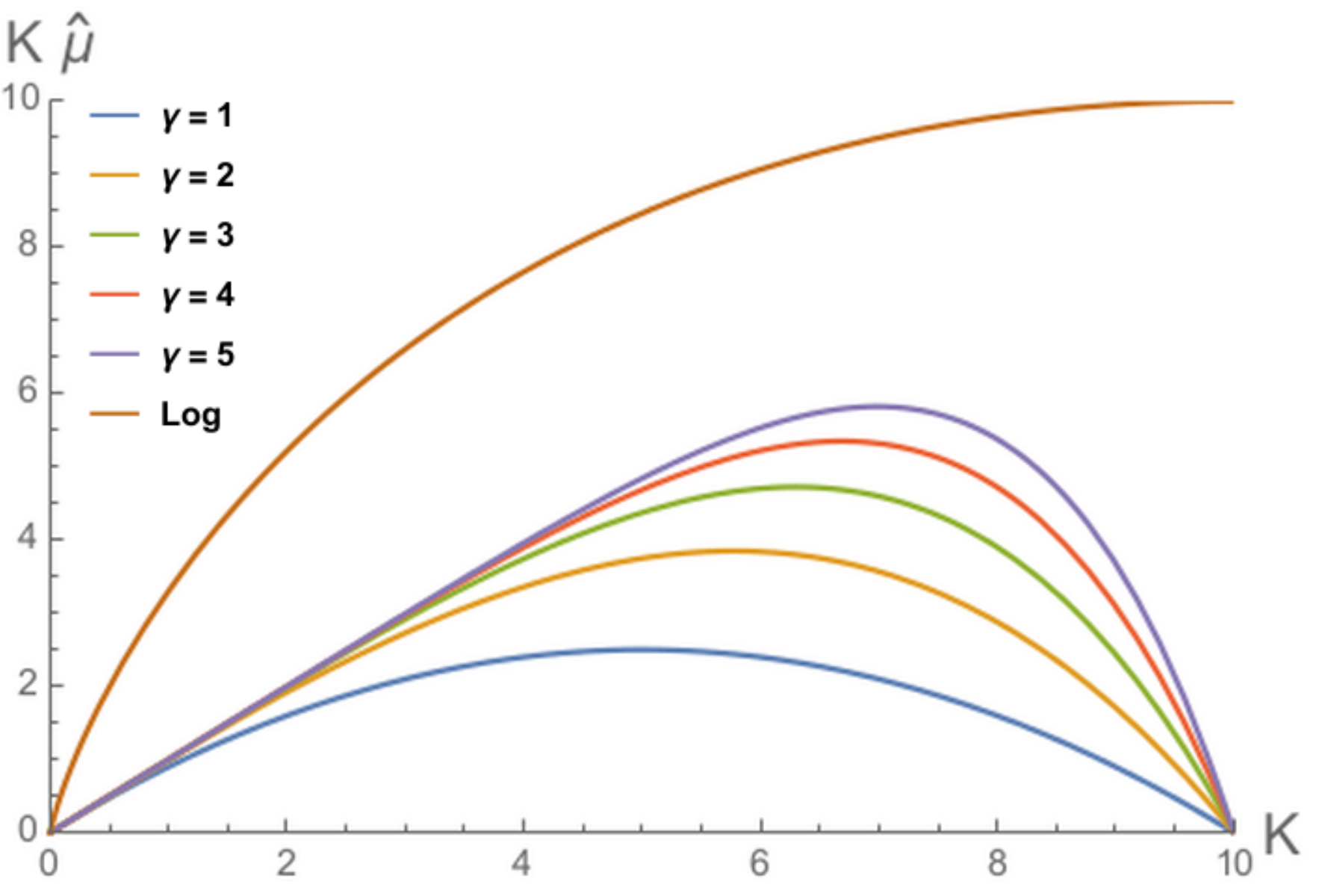}
\caption{Return functional $K \hat \mu(K)$ of eqn. (\ref{eqn:marketimpact}), for $f(x)=x^\gamma$, with $\gamma=1,...,5$, and $f(x)=\log(x)$.  The other constants are fixed at $\mu=\rho=1$, $\tilde K=10$, $\alpha = 1$.}
\label{fig:returnfunc}
\end{figure}

\
\subsection{Analytical solution for $f(x)=x^\gamma$}  \label{sec:solxgamm}
In this section we solve the stochastic equation \eqref{eq:capitalEvolution} in the case where $f(x)=x^{\gamma}$ and $\rho(t)=\rho$ is constant in time. 
We also use the simplifying assumption that the parameters $\mu$, $\sigma$ and $\gamma$ are constant in time. In this case we have that
\begin{equation}
\hat\mu(\rho K,t)= \mu\left(1-  \left(\frac{\rho K}{\tilde K}\right)^\gamma\right),
\end{equation}
where from now on we suppress the argument $t$ for the capital $K$.
If we take $\rho=\tilde K=1$ we can write
\begin{equation}
dK = \mu K (1- K^\gamma) dt + \sigma K dW_t.
\label{eq:tosolve1}
\end{equation}

Eqn. (\ref{eq:tosolve1}) can be solved using standard methods \cite{KloedenPlaten}. Here we report only the solution for $\gamma=1$, but a full derivation for generic $\gamma$ can be found in Appendix \ref{app:aa}:

{\small
\begin{eqnarray}\label{solutionLinear}
K(t)&=&\frac{\tilde K}{\rho}\ e^{ (\rho \mu-\frac{(\rho \sigma)^2}{2})t+\rho \sigma W_t }\\
\nonumber&\times& \left(\frac{\tilde K}{\rho K_0}+\rho \mu\int_0^t e^{ (\rho \mu-\frac{(\rho \sigma)^2}{2})s+\rho\sigma W_s } ds\right)^{-1}.
\label{eq:thesolution1}
\end{eqnarray}}

%\textcolor{blue}{Add comment for referee here}
The asymptotic stochastic equilibrium of the above solution can be determined by solving 
\begin{equation}
\langle d \log(K(t)/K_0) \rangle = 0.
\label{eq:stocheq}
\end{equation} 
In fact, from eqn. (\ref{eq:capitalEvolution}) and using It\^o theorem, we obtain
\begin{equation}
0=\langle d \log (K/K_0) \rangle=\rho(t) \mu \left(1-f\left(\frac{\rho(t) K(t)}{\tilde K}\right)\right)-\frac{\sigma^2}{2} \rho^2 
\end{equation}
and for linear functions $f(\cdot)$ we obtain
$K =  \tilde K(\frac{1}{\rho}-\frac{\sigma^2}{2\mu})=\tilde K(\frac{1}{\rho}-\frac{1}{2\rho_{Thorp}})$.
If the parameter $\rho$ is constant, the above implies an asymptotic equilibrium state $K(\infty)$. This equilibrium point is compatible with our simulations presented in the following.. This also holds in the case of the geometrical Brownian motion, where $\rho_{opt}\equiv \rho_{Thorp}=\frac{\mu}{\sigma^2}$ was computed in \cite{Thorp2006, Peters2}.

\subsection{Analytical solution for $f(x) = \alpha \log(x)$} \label{sec:sollogx}
In this section we provide a solution for the logarithmic functional form of carrying capacity, introduced recently in \cite{Marketimpact2} for the case of stock markets and describing cell growth. The equation of interest is
\begin{equation}
dK= \mu K\left(1-\alpha \log\left(\frac{K}{\tilde K}\right)\right) dt+ \sigma\ K\ dW_t,
\end{equation}
for the case where $\rho$ is constant. The solution is obtained in Appendix \ref{app:ab}, where we find:
\begin{widetext}
\begin{eqnarray}
K(t)&=&\exp\left[ e^{ -\alpha \mu t }\left(C_0+\left(\mu(1+\alpha \log(\tilde K))-\frac{\sigma^2}{2}\right) \frac{e^{ \alpha \mu t }-1}{\alpha \mu}+\sigma \int_0^t e^{ \alpha \mu s}dW_s\right)\right] 
\label{eq:impactlog}
\end{eqnarray}
\end{widetext}
The expectation $\mean{\log(K(t))}$ can then be evaluated analytically, using $\mean{\int_0^t f(s)dW_s}=0$ for any deterministic and continuous function $f(s)$, and thus, substituting $C_0=\log(K_0)$,
\begin{widetext}
\begin{eqnarray}
\mean{\log(K(t))}&=& e^{- \alpha \mu t} \left(C_0+\left(\left(\frac{1}{\alpha}+\log(\tilde K)\right)-\frac{\sigma^2}{2\alpha \mu}\right)(e^{\alpha \mu t}-1)\right). \nonumber \\
\label{eqn:logav}
\end{eqnarray}
\end{widetext}

Looking at the asymptotic stochastic equilibrium point again, we have
\begin{eqnarray}
\mean{\log(K(t\rightarrow\infty))}&=&\log( \tilde K)+{\frac{1}{\alpha}\left(1-\frac{\sigma^2}{2\mu}\right)}. \nonumber \\
\end{eqnarray}
We note that while the expected return and the exponential of expected log-return are not equal, they converge asymptotically.

\section{Optimal trajectories}

We now turn to the problem of finding the optimal trajectory $\rho(t)$ following the Kelly strategy that maximizes the expected log-return of the investor's capital, i.e., $\mean{\log(K(t)/K_0)} $.
In general, we must solve the equation
{\small
\begin{eqnarray}
d\log(\frac{K}{K_0})&=& \left[\mu \rho(t) \left(1-f\left( \frac{\rho(t)K(t)}{\tilde K}\right)\right) -\rho^2(t)\frac{\sigma^2}{2}\right ]dt \nonumber \\
&+& \rho(t) \sigma dW_t,
\end{eqnarray}}
which implies
{\small
\begin{eqnarray}
\mean{d\log(\frac{K}{K_0})}&=& \mean{\mu \rho(t) \left(1-f\left(\frac{\rho(t) K(t)}{\tilde K}\right)\right) -\rho^2(t)\frac{\sigma^2}{2} }dt\nonumber \\
&+&\mean{\rho(t) \sigma dW_t}.
\label{dlog}
\end{eqnarray}}

%We will later analyze the general problem in a simple situation in which $\rho(t) = \rho$. 
It is instructive to discuss first the case of a time-dependent $\rho$ in eqn. (\ref{dlog}), and then provide approximations for which we can obtain an explicit solution for the optimal parameter $\rho$.
Using \eqref{dlog}, the expected logarithmic return over a time horizon $T$ is:
{\small
\begin{align}
&\log\left(\frac{K(T)}{K_0}\right) = \nonumber\\
&\int_0^T \mean{ \mu \rho(t) \left(1-f\left( \frac{\rho(t) K(t)}{\tilde K}\right)\right) -\rho^2(t)\frac{\sigma^2}{2} }dt.
\label{eq:max}
\end{align}}
The main difference with respect to the case of no carrying capacity is the fact that due to the effective dependence of the drift term on $K$, optimizing eqn. (\ref{eq:max}) requires taking a functional derivative with respect to $\rho(t)$ and setting it to zero, i.e., 
\be\frac{\delta}{\delta \rho(t)} \log\left(\frac{K(T)}{K_0}\right)=0\,.\ee In the case of a pure geometric Brownian motion, it can be shown that the optimal $\rho(t)$ is constant in time.  In the case under consideration one strategy is to first obtain a solution for arbitrary $\rho(t)$, and then take the functional derivative in the integral of eqn. $(\ref{eq:max})$. Unfortunately, this is difficult, so in the following we will resort to two different approximations.  First we take the stationary case, in which the solution for \textit{constant} $\rho$ is known, as discussed in Sec. \ref{sec:adiabatic}. For the second approximation, if we write the function of the carrying capacity as a series expansion, we can write a dynamical set of equations for the derivatives of all the moments of the solution $\mean{K(t)^n}$ which then can be optimized iteratively; this procedure will be discussed in Sec. \ref{sec:moments}.

\subsection{Stationary approximation} \label{sec:adiabatic}
%\textcolor{red}{note that the referee said we need to change the stationary word - and I agree}
In this section we will use the exact solutions obtained in Sec. \ref{sec:solxgamm} and Sec. \ref{sec:sollogx} under the assumption of constant parameter $\rho$ within a quasi-stationary approximation scheme. More precisely, we assume that $\rho(t)$ is a slow variable with respect to $K(t)$, and that the latter quickly relaxes to what would be its asymptotic value should $\rho$ remain constant.  The validity of the approximation can then be assessed from the obtained solution by checking whether $|\mean{K(t)} \frac{\partial_t \rho(t) }{\partial_t \mean{K(t)}}| \ll 1$.

In general, we have that
{\footnotesize
\begin{eqnarray}
&&\frac{\delta}{\delta \rho(t)}\log\left(\frac{K(T)}{K_0}\right) \nonumber \\
&=&\frac{\delta}{\delta \rho(t)} \int_0^T\left [\mu \rho(t) \left(1-\mean{f\left(\rho(t) \frac{K(t)}{\tilde K}\right)}\right) -\rho^2(t)\frac{\sigma^2}{2} \right]dt \nonumber \\
&=&\int_0^T \left[ \mu \left(1-f\left(\rho(t) \frac{K(t)}{\tilde K}\right)\right) -2 \rho(t)\frac{\sigma^2}{2}\right.  \nonumber \\
 &-&\left.\rho(t) \mu \frac{\delta}{\delta \rho(t)}\mean{f\left(\rho(t) \frac{K(t)}{\tilde K}\right) }\right]dt. 
\end{eqnarray} 
} 
To find an optimal solution, we now impose the following condition:
\bea
&\mu& \left(1-f\left(\rho(t) \frac{K(t)}{\tilde K}\right)\right) -2 \rho(t)\frac{\sigma^2}{2}\\
\nonumber&-&\rho(t) \mu \frac{\delta}{\delta \rho(t)}\mean{f\left(\rho(t) \frac{K(t)}{\tilde K}\right) }=0,
\eea
where the last functional derivative requires knowledge of $K(\rho(t),t)$ for arbitrary $\rho(t)$. In order to evaluate this functional derivative, as a first approximation, we will assume that $\rho(t)\approx\rho$ in the interval $\Theta=[t,t+\delta t]$.
If this is true, then we can evolve in the interval $\Theta$ the solution with a constant $\rho$ from $t_0=t$ to $t_f=t+\delta t$ with initial condition $K_0=K(t)$. Let us call such a solution $\mathcal K(\delta t,\rho,K(t))$, which satisfies the property
$\lim_{\delta t\rightarrow 0} \mathcal K(\delta t,\rho,K(t))= K(t)$. The underlying assumption of this approach is that $\rho(t)$ changes slowly with respect to the stochastic dynamics, which implies that $|\mean{K(t)} \frac{\partial_t \rho(t) }{\partial_t \mean{K(t)}}|\ll 1$. 
Within this approximation, we can write the functional derivative as:
{\footnotesize
\begin{equation}
\frac{\delta}{\delta \rho(t)}\mean{f\left(\rho(t) \frac{K(t)}{\tilde K}\right) }\approx \partial_\rho \mean{f\left(\rho \frac{\mathcal K(t+\delta t,\rho,K(t))}{\tilde K}\right)},
\end{equation}
} 
Then the optimal instantaneous parameter $\rho(t)$ can be obtained by solving the following equation:
\begin{widetext}
\begin{equation}
 \mu \left(1-f\left(\rho(t) \frac{K(t)}{\tilde K}\right)\right) -2 \rho(t) \frac{\sigma^2}{2}=\lim_{\delta t\rightarrow 0} \rho(t) \mu\ \partial_\rho \mean{f\left(\rho(t) \frac{\mathcal K(\delta t,\rho,K(t))}{\tilde K}\right)},
\end{equation}
\end{widetext}
which is the approximation we use in the following.

\subsubsection{$f(x)=x^\gamma$: Expansion assuming $\frac{K}{\tilde K}\ll1$}

\begin{figure}
\includegraphics[scale=0.5]{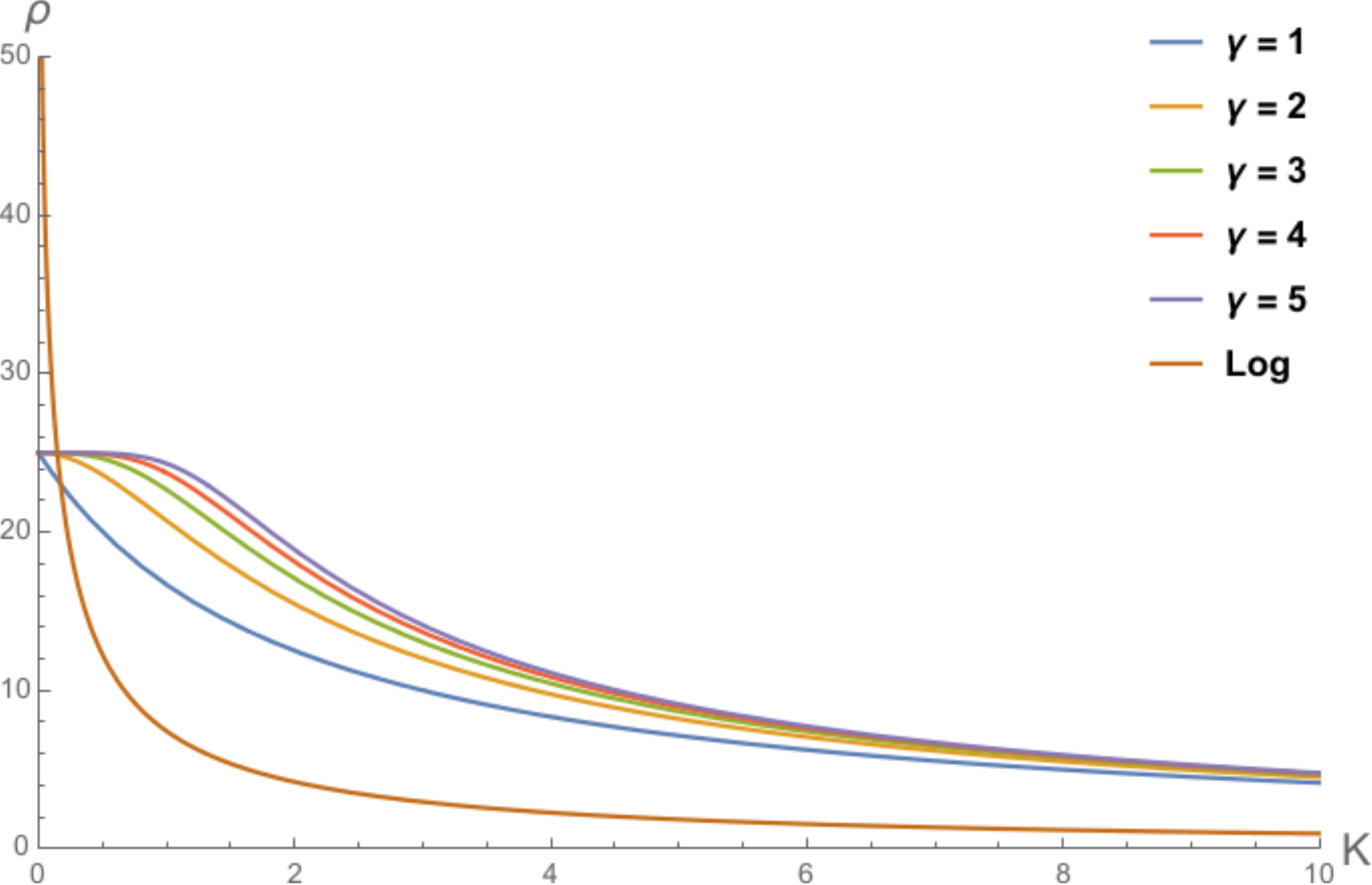}
\caption{Optimal parameter $\rho(t)$ obtained numerically from eqn. (\ref{eq:numrho}) as a function of $K\in[0,10]$, for various values of $\gamma$, and for logarithmic carrying capacity with $\alpha=1$. Other constants are fixed at $\mu=1$, $\sigma=0.2$, $\tilde K=50$.}
\label{fig:gammak}
\end{figure}

In the case of $f(x)=x^\gamma$, evaluating $\mean{f}$ is a non trivial task. Even assuming we have the stationary approximation,  expanding equation \eqref{solution} in $K/\tilde K$, and considering only the zeroth order term of this expansion, we obtain the following expression to be solved for $\rho$,
\begin{equation}
\mu  \left(1-\left(\frac{K \rho }{\tilde{K}}\right)^{\gamma }\right)-\rho  \sigma ^2=0,
\label{eq:numrho}
\end{equation}
which cannot be solved analytically for arbitrary values of $\gamma$. 
However, for $K\gg\tilde K$, we can obtain the approximate solution
\begin{equation}
\rho(K\gg \tilde K)\approx \frac{\tilde{K}}{K},
\end{equation}
which is independent of $\gamma$. A plot with the numerical solutions of $\rho(K)$ obtained from \eqref{eq:numrho} for different values of $\gamma$ is shown in Fig. \ref{fig:gammak}.

In the particular case of $\gamma=1$ the solution is simply:
\begin{equation}
\rho(t)=\frac{\mu}{2 \mu \frac{K(t)}{\tilde K} +\sigma^2},
\label{eq:optimalrho}
\end{equation}
which shows explicitly that the presence of carrying capacity effectively increases the risk, as the optimal fraction of resources to be deployed is a decreasing function of $K$.\footnote{Such calculation can be repeated in the presence of a risk-free asset with return $\mu_{rf}$. In this case, $\mu_{rf}$ would simply be added at the denominator of eqn. (\ref{eq:optimalrho}).} This is important as is it a generalization of the result of \cite{Peters1} and has direct applications to the problem of optimal trajectories in the context of financial time series where one has an embedded transaction cost. This applies for instance to lotteries and wholesale electricity markets, where one has independent processes at each time step. In this case, the transaction cost plays the role of market impact.

Surprisingly, the optimal parameter of eqn. (\ref{eq:optimalrho}) holds up to second order in $\xi=\frac{K}{\tilde K}$. In order to obtain precise estimates of the parameter $\rho$, the expectation $\mean{K}$ must be evaluated. This is done up to second order in $\xi$ in Appendix \ref{sec:appendix}, using techniques partly developed in \cite{Caravellietal}.

In order to check the consistency of the stationarity approximation, we evaluate 
\begin{equation}
\partial_t \rho(t)=\partial_t \frac{\mu}{2 \frac{K(t)}{\tilde K}+\sigma^2}= -\frac{\mu}{(2 \frac{K(t)}{\tilde K}+\sigma^2)^2} \left(2 \frac{\partial_t K(t)}{\tilde K}\right),
\end{equation}
which implies
\begin{equation}
\left|\frac{\partial_t \rho(t)}{\frac{\partial_t K(t)}{K(t)}}\right|= \frac{2\mu}{\left(2 \frac{K(t)}{\tilde K}+\sigma^2\right)^2}  \frac{K(t)}{\tilde K}.
\label{eq:small}
\end{equation}
The right hand side of eqn. (\ref{eq:small}) is small as long as $\frac{K(t)}{\tilde K}\ll 1$, consistent with the expansion we performed.

Next we take the optimal $\rho(t)$ from above, and study the implied stochastic differential equation for $K(t)$.  For the case of $\gamma=1$ we have:
\begin{equation}
d \log(K/K_0)=\frac{1}{2} \frac{\mu^2}{2 \mu \frac{K}{\tilde K}+\sigma^2} dt +\frac{\mu \sigma}{2 \mu \frac{K}{\tilde K}+\sigma^2} dW.
\label{eq:optdiffrho}
\end{equation}
When $K\gg \frac{\sigma^2 \tilde K}{\mu}$, this simplifies to:
\begin{equation}
\langle d\log(K/K_0) \rangle=\frac{1}{4} \frac{\mu \tilde K}{K} dt.
\label{eq:lintheory}
\end{equation}
Since we observe that the asymptotic growth is compatible with a linear function of $K$, we can obtain the proportionality constant by using the ansatz $K(T)\approx a T$, we obtain that the slope of the linear approximation is $\frac{\mu \tilde K}{4}$ for $T\gg \frac{\sigma^2}{4 \mu^2}$.  Further, when $\tilde K \rightarrow \infty$, this slope $\rightarrow \infty$ as well, because we are asymptotically approximating an exponential with a linear function.

\subsubsection{$f(x)= \alpha \log(x)$} 

For the case of a logarithmic carrying capacity term we have shown how to evaluate explicitly $\mean{K(t)}$ and $\mean{\log(K(t))}$ in eqns. (\ref{eqn:logav}). Using this solution, if we assume  the stationary approximation in which $\rho$ changes slowly compared to $K$, we obtain the following equation for the optimal parameter $\rho$ in the limit $\delta t\to 0$:
\begin{equation}
\alpha  \mu  \left(\log
   \left(\frac{\tilde K }{\rho }\right)-\log
   (\text{K(t)}/K_0)\right)+\mu(1-\alpha) -\rho  \sigma ^2=0,
\end{equation}
from which we can solve for $\rho(t)$:
\begin{equation}
\rho_{opt}(t)\approx \alpha W\left(\frac{K_0 e^{\frac{1}{\alpha }-1}
   \tilde K \sigma ^2}{\alpha  K(t) \mu }\right)  \frac{  \mu  }{\sigma ^2},
   \label{eqn:optlog}
\end{equation}
where $W$ is the Lambert W-function. Note that for $\alpha\rightarrow 0$ we recover again the result of \cite{Peters1}. This allows us to evaluate the critical ratio $ \xi=\frac{K}{\tilde K}$ for which $\rho_{opt}=1$, which is given by $\xi=e^{\frac{1}{\alpha}(\frac{\sigma^2}{\mu}-1)-1}$. 

Similarly to the case of a linear carrying capacity given in eqn. (\ref{eq:optdiffrho}), we can obtain an effective differential equation by inserting the obtained optimal $\rho(t)$ of eqn. (\ref{eqn:optlog}). Using the asymptotic properties of the Lambert W-function in the limit $K\gg 1$, this differential equation is given by 
\begin{equation}
\mean{d \log(K)}\approx \tilde K \mu dt,
\label{eq:lintheorylog}
\end{equation}
which implies an asymptotic linear growth given by $K(T)\approx  4 a T$, where $a=\frac{\mu \tilde K}{4}$ is the slope obtained in eqn. (\ref{eq:lintheory}). We thus have the result that in the case of logarithmic carrying capacity, the Kelly strategy implies a growth rate which is asymptotically twice the rate obtained for a linear function.

\subsection{Analytical result: short timescales} \label{sec:moments}

The approach of the previous section has some drawbacks, in particular, evaluating cumulants of the exponential of the Brownian motion is a lengthy task in general. While for the case of logarithmic carrying capacity it is possible to evaluate the averages exactly, this is not true in the general case. 
As an alternative, we can proceed by directly integrating the equations for the moments and
elaborating an approximation scheme based on the smallness of the time horizon
with respect to the other scales.

It is reasonable to expect that at least for small variations the carrying capacity can be parametrized
with a power series, leading to the following stochastic differential equation:
\begin{equation}
dK = \rho \mu K \left(1+ 
\sum_{k=1}^{n} \lambda_k (\rho K)^{k \gamma}
\right)dt
+ \sigma \rho K dW.
\end{equation}
As before, we focus on the maximization of the expectation value of the logarithm of $K$. Using Ito's Lemma,
\begin{equation}
\mean{\frac{d}{dt} \log K/K_0} = \mu \rho \left(
1+ {\sum_{k=1}^{n} \lambda_k \rho^{k \gamma} \mean{K^{k \gamma}}} 
\right) -\frac{\sigma^2 \rho^2}{2}\, .
\end{equation}
To solve this equation for generic values of the parameters $\lambda_{k}$,
we need to compute all the moments $\langle K^{m\gamma}\rangle$, that is,
we need to solve the equations of motion for these observables:
\begin{align}
&\mean{\frac{d}{dt}K^{m\gamma}}
= \mu \rho \gamma m \mean{ \left( 1 +\sum_{k=1}^{n} \lambda_k \rho^{k\gamma} K^{k \gamma} \right) K^{m\gamma}} \nonumber \\
&+\ \gamma m (\gamma m -1) \frac{\sigma^2 \rho^2}{2} \mean{K^{m \gamma}}.
\end{align}
These form a tower of coupled equations\footnote{Formally, this tower can be rewritten in the form:
$$\dot{\vec e}=M \vec e, $$
with infinite dimensional objects.
A formal solution, for given (time independent) $M$, 
is $e_m(t)=\sum_k R_{mk}(t) e_k(0)$, with $R$ being the exponential of the operator
$M$, $R(t)=e^{M t}$. }.  Using the notation $e_{m} := \mean{K^{m\gamma}}$, we can write the above as:
\begin{equation}
\dot{e}_{m} = \mu \rho \gamma m \sum_{k=0}^{n} \lambda_{k} \rho^{k\gamma} e_{m+k} +\\
\gamma m (\gamma m -1) \frac{\sigma^2\rho^2}{2} e_{m},
\end{equation}
where
$
\lambda_0 = 1
$. 
The initial conditions are $e_{m}(t_0) = K_0^{\gamma m}$, as the PDF for $K(t=t_0)$ is a
Dirac delta at the initial time.

With these equations, given a time horizon $\delta t$, we can compute the 
Taylor expansion of the derivative $d\log K(t)/dt$ at any order in an expansion in $\delta t$.  In general, this is given by:
\begin{equation}
\frac{d}{dt} \log(K(t_0+\delta t)) = \Phi(K_0,\mu,\sigma,\gamma,\{\lambda_k\};\delta t),
\end{equation}
with
\begin{equation}
 \Phi(K_0,\mu,\sigma,\gamma,\{\lambda_k=\delta_k^0\};\delta t) =  \mu \rho - \frac{{ \sigma}^2\rho^2}{2}.
\end{equation}

Using this general procedure, the maximisation needed to determine $\rho$ is straightforward, once a truncation in the expansion in $\delta t $ has been fixed.
As an example, consider the case in which \begin{align}
\gamma=1, \qquad n=1,
\qquad \lambda_1 = - \frac{1}{ \tilde K}. 
\end{align}
To first order in $\delta t$ we then have:
\begin{eqnarray}
\mean{\frac{d}{dt} \log K} \simeq\mu  \rho -\frac{\rho ^2 \sigma ^2}{2} + \nonumber\\
 -
  \frac{{K_0} \mu  \rho ^2}{\tilde K}+\left(
 \frac{K_0^2 \mu ^2 \rho ^4}{{\tilde K}^2}-\frac{ K_0 \mu ^2 \rho
   ^3}{{\tilde K}}\right) \delta t \,
  .
\end{eqnarray}
As expected, the corrections to the geometric Brownian motion case are controlled by the quantity  ${\frac{K_0}{\tilde K} }$.
With respect to these quantities, the optimal parameter  $\rho$ is:
\begin{equation}
\rho_{opt} \simeq
\frac{ \mu }{2 \mu  \frac{K_0}{\tilde K}+ \sigma ^2}+\frac{ \left(-3 {\tilde K}^3 \mu ^4 \sigma ^2 K_0-2 {\tilde K}^2 \mu ^5 K_0^2\right)}{\left(2 \mu  K_0+\tilde K \sigma
   ^2\right)^4} \delta t.
   \label{eqn:eqnfirstorder}
   \end{equation}
In this last expression we recognize, at the zeroth order, the same term obtained in the approximation in which $\rho(t)$ is constant. At the first order we obtain a correction proportional to the size of the time horizon $\delta t$.
These results can be generalized without difficulty to higher orders. 

What is remarkable in the result of this short timescale analysis is that the optimal value for $\rho$, in the general case, is a function of the initial condition,
the parameters of the process \emph{and} of the time horizon.

\subsection{Numerical simulations}

In this section we present numerical tests of our analytical results, using a Monte Carlo approach and a stochastic Euler method to solve the differential equations.

\begin{figure}
\includegraphics[scale=0.355]{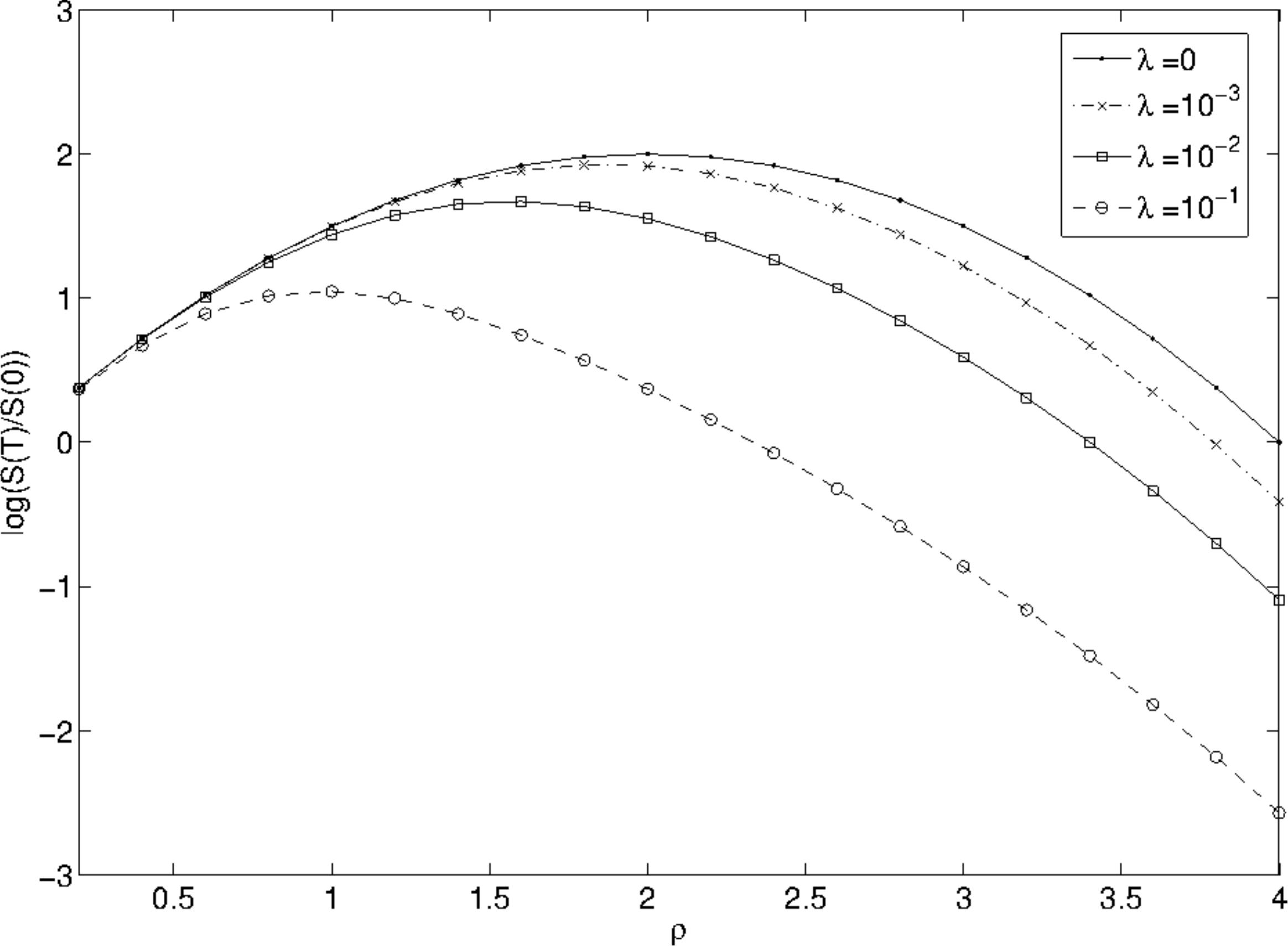}
\caption{ Log return as a function of $\rho$ at $T=1$ for parameters $K(0)=1$, $\mu = 2$, $\sigma=1$, $\gamma = 1$ 
and various values of $\lambda = 1/\tilde{K}$.} 
\label{fig:curves}
\end{figure}

First, In Fig. \ref{fig:curves} we report the expectation values of log returns over a fixed time horizon $ T =1$ and for different values of the parameter $\rho$, assumed to be constant in time, and for different values for the scale of carrying capacity. Each point is an estimate for the expectation of the endpoint of the numerical solution of the corresponding stochastic differential equation. 
The curves that are obtained from these points clarify how the log returns reach their maximum for special choices of $\rho$.
The points on the upper curve are obtained in the case of no dependence on the amount of resources, i.e., the simple geometric Brownian motion. They match the results of \cite{Peters1} for the values of the parameters that we are considering. As expected, the plot shows that the optimal $\rho$ decreases with the strength of carrying capacity. 

\begin{figure}
\includegraphics[scale=0.35]{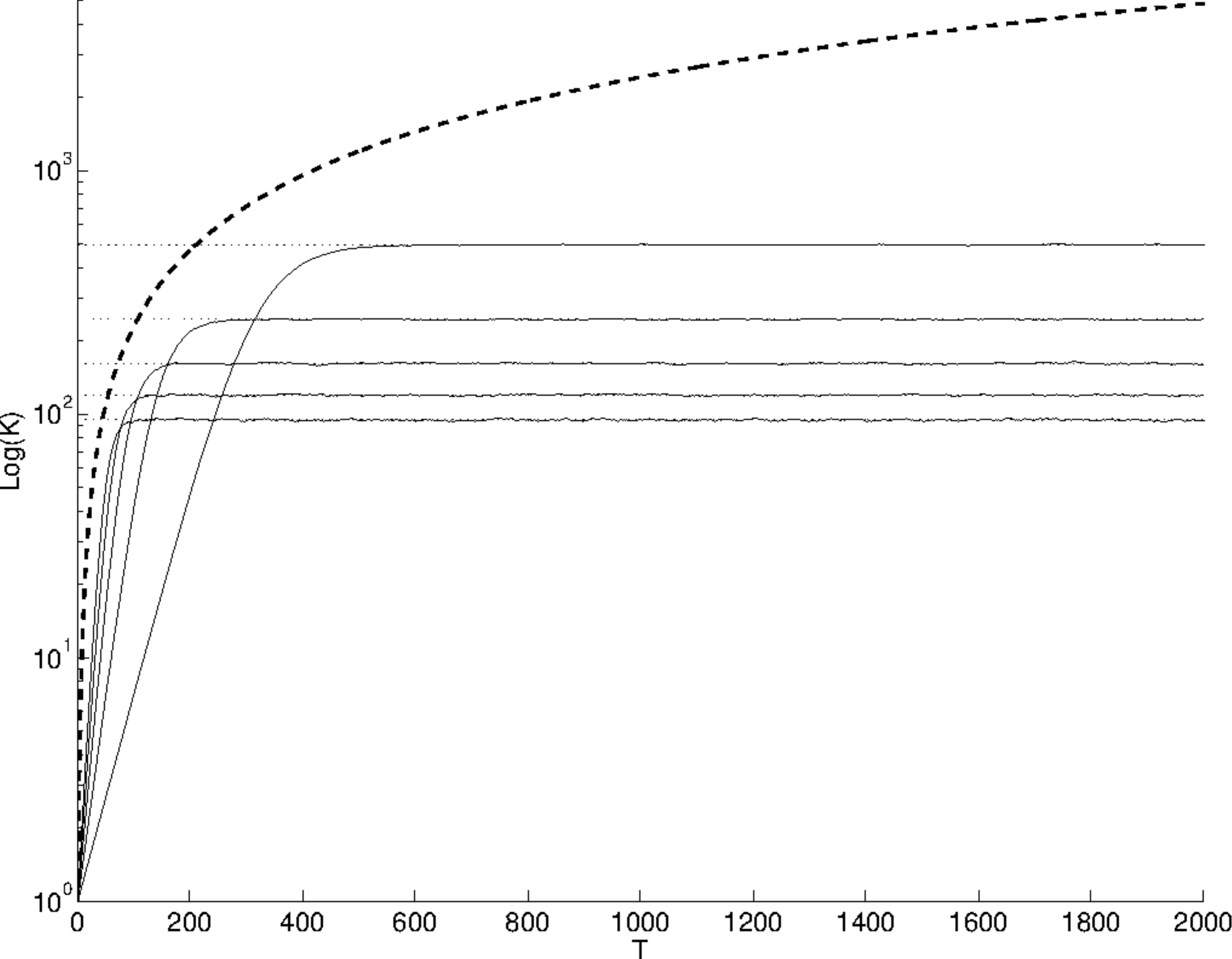}
\caption{Optimal Kelly path for $K$ with carrying capacity function $x^\gamma$ plotted against the case of those fixed  at $\rho=1,0.8,0.6,0.4,$ and $0.2$ (full lines).The parameters used are $K(0)=1$, $dt=0.01$, $\mu=0.1$, $\sigma=0.1$, $\tilde K=10$ and $\gamma=1$, averaged over $2000$ samples. The horizontal dashed lines represents the stochastic equilibria obtained from the analytic formula.}
\label{fig:kelly}
\end{figure}

\begin{figure}
\includegraphics[scale=0.35]{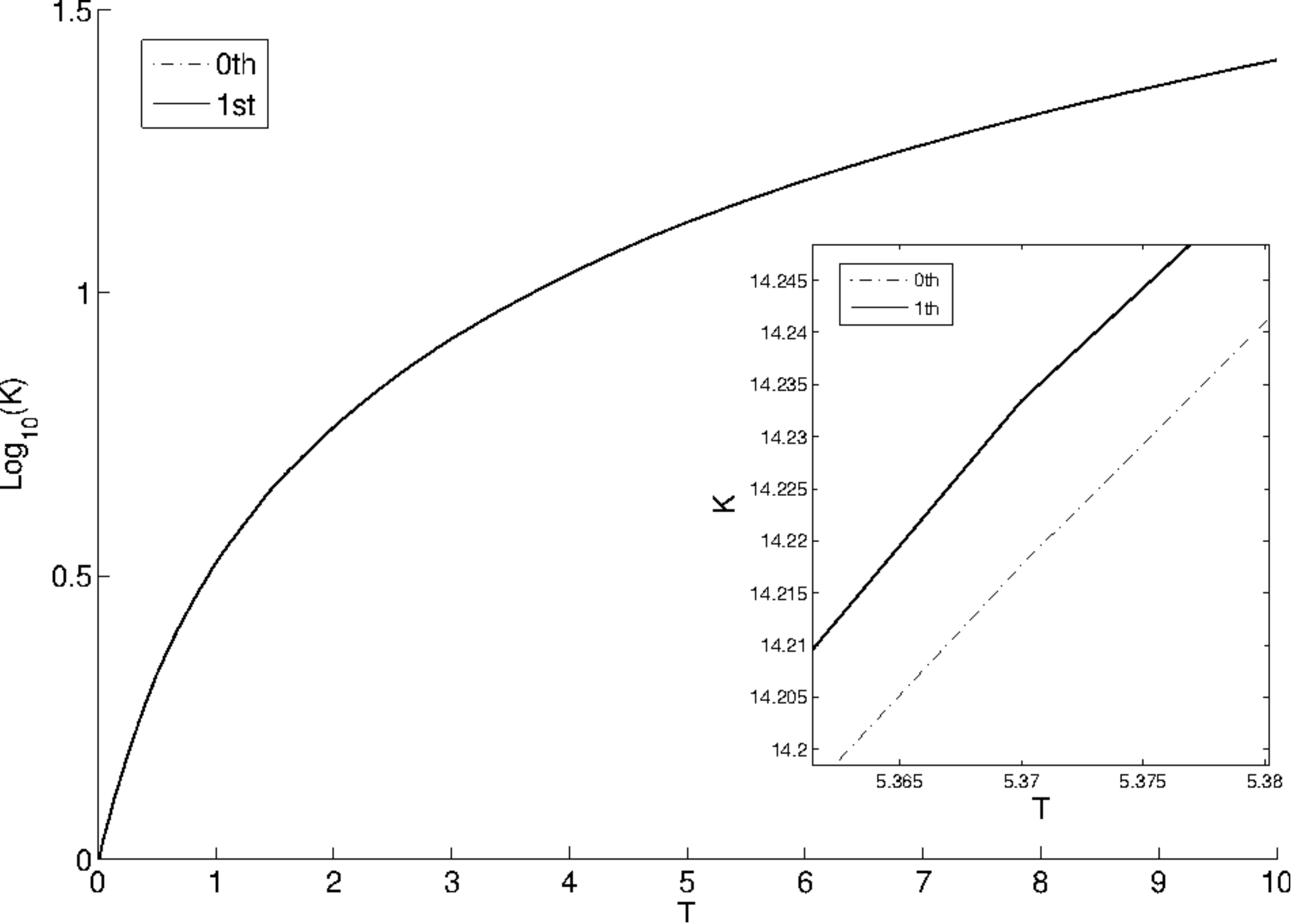}
\caption{Expected value of $K$ for the case of power law carrying capacity, with the optimal parameter $\rho$ evaluated using the 0th order in the time horizon (dashed) and 1st (solid) order correction, for $K_0=1$, $\tilde K=100$, $dt=0.01$, $\sigma=0.2$, $\mu=1$ and $\gamma=1$.  In order to distinguish the two curves, we averaged over 1000 Monte Carlo runs. We observe that the solution obtained at the first order outperforms the one obtained at the zeroth order.}
\label{fig:kellyfirst}
\end{figure}

In Fig. \ref{fig:kelly} we use eqn. (\ref{eqn:eqnfirstorder}) at zeroth order and compare to the strategy at constant $\rho=1$. We can see that the dynamic strategy outperforms those which are kept constant, and that it works reasonably well also in the case in which $K(t)\approx \tilde K$. In Fig. \ref{fig:kellyfirst} we compare this result with the first order approximation in $\delta t$. In the inset we see that the latter outperforms the optimal solution obtained at zeroth order, although the difference between the two is overall relatively small.  For $\gamma \neq 1$, our solution also outperforms the constant solution.

\begin{figure}
\includegraphics[scale=0.33]{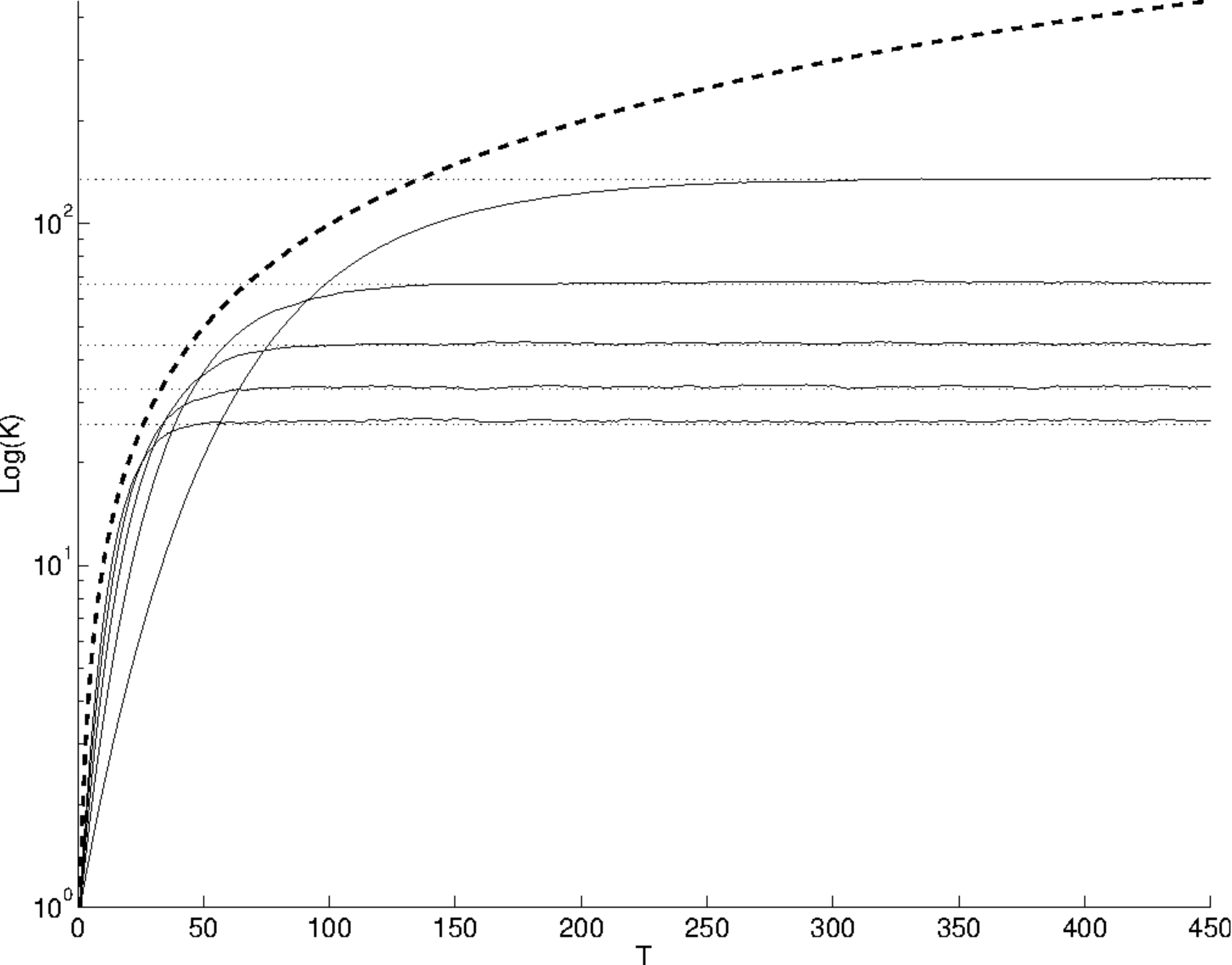}
\caption{Optimal Kelly time path of $K$ for logarithmic carrying capacity compared to the case of those fixed at $\rho=1,0.8,0.6,0.4,$ and $0.2$ (full lines).The parameters used were $K(0)=1$, $dt=0.01$, $\mu=0.1$, $\sigma=0.1$, $\tilde K=10$ and $\gamma=1$, averaged over $2000$ samples. The horizontal dashed lines represents the stochastic equilibria obtained from the analytic formula.}
\label{fig:kellylog}
\end{figure}

For logarithmic carrying capacity, we can see in Fig. \ref{fig:kellylog} that our optimal parameter $\rho(t)$ solution again outperforms the constant solution. Also note that the stochastic equilibrium obtained from eqn. (\ref{eq:stocheq}) is confirmed both in Fig. \ref{fig:kelly} and Fig. \ref{fig:kellylog}.

\begin{figure}
\includegraphics[scale=0.355]{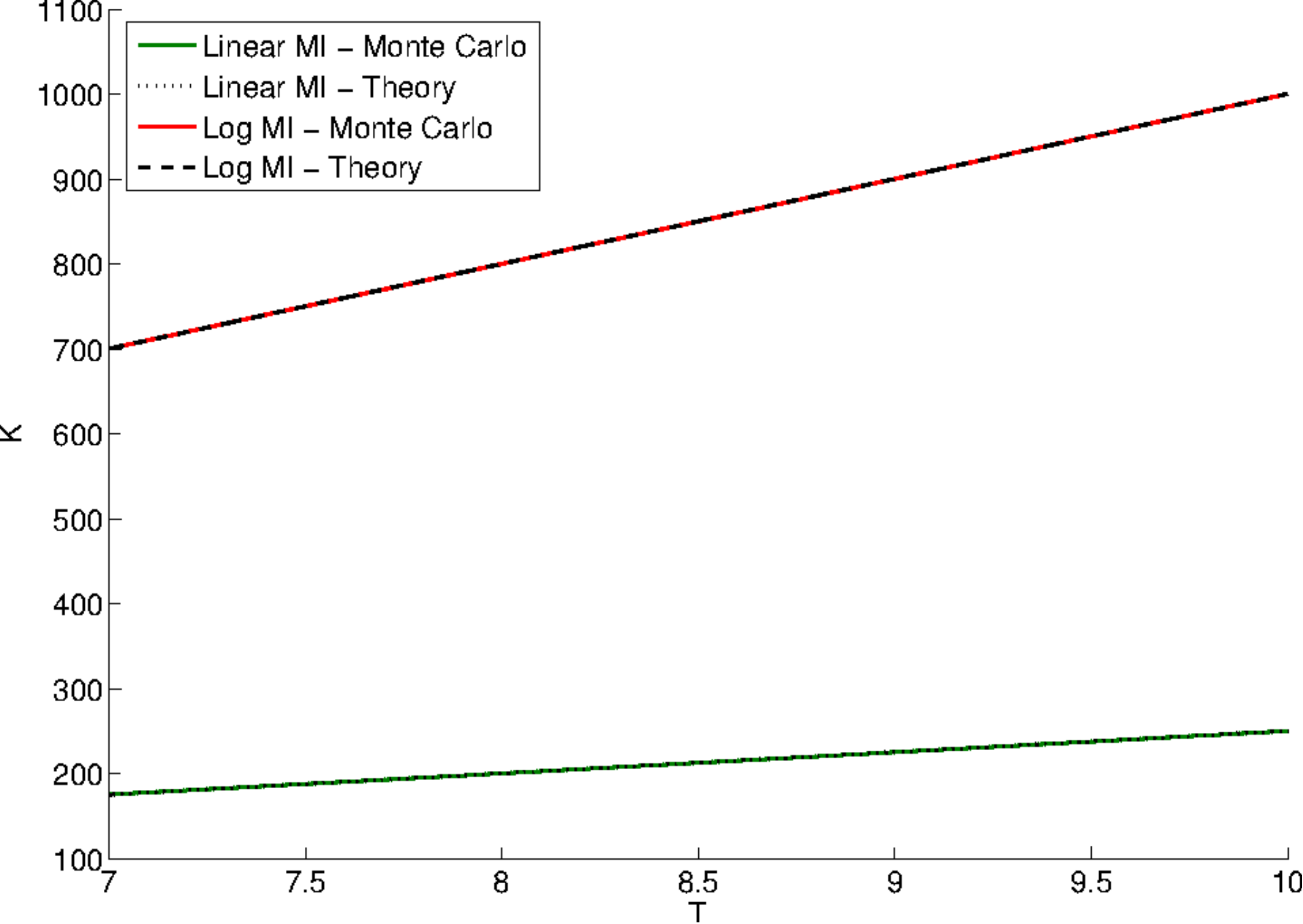}
\caption{Average asymptotic  ($T\approx 7$) linear regime in the case of linear (green curve) and logarithmic (red curve) carrying capacity for the case of $\sigma=0.2$, $\mu=1$ and $\tilde K=10$ obtained using a Monte Carlo averaged over 2000 samples, with integration step $dt=0.01$. The dashed lines represent the comparison with the linear coefficients obtained from theory, eqns. (\ref{eq:lintheory}) and (\ref{eq:lintheorylog}). } 
\label{fig:linearregime}
\end{figure}

Finally in Fig. \ref{fig:linearregime} we compare the linear regimes for the case of linear and logarithmic carrying capacity obtained in eqns. (\ref{eq:lintheory}) and (\ref{eq:lintheorylog}) to the curves obtained with Monte-Carlo simulations, showing that the slopes obtained analytically are a good match with the numerical ones.

\section{Conclusions}

In this paper we computed optimal strategies for the problem of maximal growth using the Kelly criterion, in the case of a drift term which represents the presence of carrying capacity.
Using two different methods, one considering exact solutions at constant leverage, and the second solving for the optimal solution at fixed time horizon, we obtained the same result at the lowest order. Our solutions were also tested numerically, confirming that these are optimal as compared to the case in which carrying capacity is ignored. 

We considered two specific carrying capacity functions for which empirical evidence has been presented in the literature \cite{Marketimpact1,Marketimpact2}: a power law and a logarithmic function. 
In the case of a power law carrying capacity function, we have shown that in order to evaluate the optimal solution various approximations have to be used, but we have shown that the zeroth order is correct up to second order in the parameter controlling the scale of the carrying capacity, and provided a solution up to first order for the case of a finite time horizon. In the case of a logarithmic carrying capacity function, expectations can be evaluated exactly.

The main difference between the case of the geometric Brownian motion and the case with carrying capacity is that the former requires a constant leverage while in the latter the leverage has to be dynamically adapted. 
The operator has to adapt his strategy continuously depending on his position with respect to the capacity parameter. Our results support this intuitive observation and, at the same time, complement it with concrete procedures for quantitative estimates.

Finally, the analysis presented heavily relies on the Gaussian nature of the noise term and on the powerful results of Ito's calculus. Our results can then be seen as a first assessment of the effects of carrying capacity on investment strategies, which, however, will require further elaboration. In particular, a natural extension will be the inclusion of more realistic noise terms, like general Levy processes. The investigation of the impact of a more detailed noise structure on the optimal leverage will be the subject of future work.

\section*{Acknowledgements}
F. Caccioli acknowledges support of the Economic and Social Research Council (ESRC) in funding the Systemic Risk Centre (ES/K002309/1). We also thank Ole Peters and Vladislav Malyshkin for comments on our first draft.

\appendix
\section{Solution of the stochastic differential equations for constant $\rho$}
\subsection{Power law function}\label{app:aa}
In this Appendix we discuss the solution of the equation:
\begin{equation}
dK = \mu K (1- K^\gamma) dt + \sigma K dW_t.
\label{eq:tosolve}
\end{equation}
Using the change of variables $y = K^{-\gamma}$ and Ito's Lemma, we have:
\begin{equation}
dy = \left( -\gamma \mu (y-1) + \frac{\sigma^2 \gamma(\gamma+1)}{2} y \right)dt
- \gamma \sigma y dW_t.
\label{eq:inlinsde}
\end{equation}
Notice that eqn. (\ref{eq:inlinsde}) is a stochastic differential equation of the form:
\begin{equation}
dz = (a z + c) dt + (b z+d) dW,
\label{eqn:easyform}
\end{equation}
where
\begin{eqnarray}
a&=& \frac{\sigma^2 \gamma(\gamma+1)}{2}-\gamma \mu\nonumber \\
b&=&-\sigma \gamma \nonumber \\
c&=& \gamma \mu \nonumber \\
d&=& 0.
\end{eqnarray}
This is an inhomogeneous linear stochastic differential equation with multiplicative
noise \cite{KloedenPlaten}, and has a known solution. If we define
\begin{eqnarray}
\Phi_t &\equiv&  \exp\left( (a-b^2/2)t + b W_t \right) \nonumber \\
\rightarrow \Phi_t  &=&\exp\left( -(\mu-\sigma^2/2)t - \sigma W_t \right),
\end{eqnarray}
the solution is then given by
\begin{equation}
z = \Phi_t \left( z_0 + (c-bd) \int_0^{t}\Phi_s^{-1}ds
+ d \int_0^{t} \Phi_s^{-1} dW_s \right).
\end{equation}
Writing
\begin{equation}
f_t(a,b,c,z_0) = \Phi_t \left( z_0 + c \int_0^{t}\Phi_s^{-1}ds
 \right),
\end{equation}
we obtain
\begin{equation}
y(t) = f_t\left(-\gamma \mu + \frac{\sigma^2\gamma(\gamma+1)}{2}, -\gamma \sigma, \gamma b,y_0\right).
\end{equation}
Going back to the variable $K(t)$, we have
\begin{equation}\label{solution}
K(t) = \left[ f_t\left(-\gamma \mu + \frac{\sigma^2\gamma(\gamma+1)}{2}, -\gamma \sigma, \gamma b,y_0\right) \right]^{-1/\gamma}.
\end{equation}

If we examine the special case of $\gamma = 1$, and insert again the constants $\rho$ and $\tilde K$ by rescaling $\mu\rightarrow \rho \mu$ and $\sigma\rightarrow \rho \sigma$, $\tilde K \rightarrow \tilde K/\rho$, we obtain the full solution in terms of all the original parameters:
{\small
\begin{eqnarray}\label{solutionLinear}
K(t)&=&\frac{\tilde K}{\rho}\ e^{ (\rho \mu-\frac{(\rho \sigma)^2}{2})t+\rho \sigma W_t }\\
\nonumber&\times& \left(\frac{\tilde K}{\rho K_0}+\rho \mu\int_0^t e^{ (\rho \mu-\frac{(\rho \sigma)^2}{2})s+\rho\sigma W_s } ds\right)^{-1},
\end{eqnarray}}
which is the solution used in this paper in the case of a power law carrying capacity.

\subsection{Logarithmic function}\label{app:ab}
In the case of a logarithmic carrying capacity function, and again if $\rho(t)=\rho = 1$, we have the following differential:
\begin{equation}
dK= \mu K\left(1-\alpha \log\left(\frac{K}{\tilde K}\right)\right) dt+ \sigma\ K\ dW_t,
\end{equation}
which is a stochastic Gompertzian-type of equation \cite{Gompertz, Ferrante2000, Lo2009}.
Such equation for instance appear also in the growth of reproducing cells, where now $\tilde K/\rho(t)$ represents the amount of nutrient accessible to the cells. This implies that there is a parallel between optimal leverage trajectories and optimal cell growth.\footnote{It is interesting to note that in general the logarithmic carrying capacity function can be thought as the asymptotic limit of a power law  function, as one has $\lim_{\alpha\rightarrow \infty} \alpha(1-x^{\frac{1}{\alpha}})=-\log(x)$. } If we change variables to $y=\log(K)$, then through Ito's Lemma the above becomes:
\begin{equation}
dy= \left[\mu \left(1-\alpha y+\alpha \log(\tilde K)\right)-\frac{\sigma^2}{2}\right] dt+ \sigma\ dW_t.
\end{equation}

This has the same form as eqn. (\ref{eqn:easyform}) in the previous section, with:
\begin{eqnarray}
a&=&-\alpha \mu \nonumber \\
b&=&0 \nonumber \\
c&=&\mu(1+\alpha \log(\tilde K))-\frac{\sigma^2}{2} \nonumber \\
d&=& \sigma.
\end{eqnarray}
We then have that
\begin{equation}
\Phi_t = \exp\left( -\alpha \mu t \right),
\end{equation}
and thus one obtains
\begin{widetext}
\begin{eqnarray}
K(t)&=&\exp\left[ e^{ -\alpha \mu t }\left(C_0+\left(\mu(1+\alpha \log(\tilde K))-\frac{\sigma^2}{2}\right)\int_0^t e^{ \alpha \mu s } ds+\sigma \int_0^t e^{ \alpha \mu s}dW_s\right)\right] \nonumber \\
&=&\exp\left[ e^{ -\alpha \mu t }\left(C_0+\left(\mu(1+\alpha \log(\tilde K))-\frac{\sigma^2}{2}\right) \frac{e^{ \alpha \mu t }-1}{\alpha \mu}+\sigma \int_0^t e^{ \alpha \mu s}dW_s\right)\right] 
\label{eq:impactlog}
\end{eqnarray}
\end{widetext}

\section{Average $\mean{K}$ for $f(x)=x$}\label{sec:appendix}

In this appendix we evaluate the average $\mean{K}$ as an expansion of $\xi=\frac{K}{\tilde K}$ of the denominator of the solution in eqn. (\ref{eq:thesolution1}), and show that the optimal leverage obtained in eqn. (\ref{eq:optimalrho}) holds up to second order in $\xi$.
For simplicity, we will set $\rho=1$ during the calculation of the averages, and then restore $\rho\neq1$ by rescaling $\sigma \rightarrow \rho \sigma$, $\mu \rightarrow \rho \mu$ and $\tilde K\rightarrow \tilde K/\rho$.
In this case, expanding \eqref{solutionLinear} to  order $\left(K(t)/\tilde K\right)^2$ we have that
{\small
\begin{eqnarray}
\mathcal K(\delta t,\rho=1,K(t)))&=& \nonumber \\
\tilde K\ e^{ (\mu-\frac{\sigma^2}{2})\delta t+\sigma W_{\delta t} }\left(\frac{\tilde K}{K(t)}+\mu\int_0^{\delta t} e^{ (\mu-\frac{\sigma^2}{2})s+\sigma W_s } ds\right)^{-1}&\approx& \nonumber \\
K(t)\ e^{ (\mu-\frac{\sigma^2}{2})\delta t+\sigma W_{\delta t} }\left(1-\frac{\mu K(t)}{\tilde K}\int_0^{\delta t} e^{ (\mu-\frac{\sigma^2}{2})s+\sigma W_s } ds\right). 
\end{eqnarray}
}
Taking the expectation of the above, we have:
\begin{equation}
\mean{\mathcal K}=\mean{K_{\tilde K=\infty}(t)}-\frac{\mu K(t)}{\tilde K}\mean{K_{\tilde K=\infty}(t)F[W,\delta t]},
\end{equation}
with $F[W,t]=\int_0^t e^{ (\mu-\frac{\sigma^2}{2})s+\sigma W_s } ds$ being the integral of an exponential gaussian process and ${\mathcal K}_{\tilde K=\infty}=K(t) e^{(\mu -\frac{\sigma^2}{2}) \delta t+\sigma W_{\delta t}}$.
Using $\mean{e^{\sigma W_s}}=e^{\frac{\sigma^2}{2} s}$, we then have that
\begin{equation}
\mean{{\mathcal K}_{\tilde K=\infty}}=K(t) e^{\mu \delta t},
\end{equation}
and further using $\mean{e^{\sigma (W_s+W_{s^\prime})}}=e^{\frac{\sigma^2(s+s^\prime+2 \text{min}(s,s^\prime))}{2}}$, we get
\begin{eqnarray}
\mean{\mathcal K_{\tilde K=\infty}(t)F[W,\delta t]}&=&K(t) e^{\mu \delta t}\int_0^{\delta t} e^{ (\mu+\sigma^2) s} ds \nonumber \\
&=&\frac{K_0}{\mu+\sigma^2} e^{\mu t}(e^{(\mu+\sigma^2) t}-1).
\end{eqnarray}
Putting these together, 
{\footnotesize
\begin{eqnarray}
\mean{\mathcal K(\delta t,\rho=1,K(t)} &=& K(t) [(1-\frac{K(t)}{\tilde K}\frac{\mu}{\mu+\sigma^2})e^{\mu \delta t}\nonumber \\
&-&\frac{K(t)}{ \tilde K}\frac{\mu}{\mu+\sigma^2}e^{(2 \mu+\sigma^2) \delta t}] +O(\frac{ K_0}{\tilde K})^2, 
\end{eqnarray}
}
which is valid in the approximation $K_0 e^{\mu t} \ll \tilde K$.  Restoring $\rho$, the above becomes
\begin{eqnarray}
&\langle&\mathcal K(\delta t,\rho,K(t)\ \rangle=K(t) [(1-\rho \frac{K(t)}{\tilde K}\frac{\mu}{\mu+\rho \sigma^2})e^{\rho \mu \delta t}\nonumber \\
&-&\rho \frac{K(t)}{ \tilde K}\frac{\mu}{\mu+\rho \sigma^2}e^{(2 \rho \mu+\rho^2 \sigma^2) \delta t}] +O\left(\frac{ K_0}{\tilde K}\right)^2.
\end{eqnarray}
We now impose $\partial_\rho [\rho(t)\mu- \rho(t)^2  ( \mu\frac{\mean{\mathcal K(\delta t,\rho,K(t)}}{\tilde K}+\frac{\sigma^2}{2})]=0$, and
after expanding at order $(K(t)/\tilde K)^2$ and imposing $\delta t\to0$, we obtain again the equation for $\rho$:
\begin{equation}
\mu -\rho  \sigma ^2-\frac{2 \mu  \rho 
   K(t)}{\tilde K}=0.   
\end{equation}
This implies that the optimal leverage obtained at zeroth order is valid up to the second order in $K(t)/\tilde K$.
In general, it is possible to use perturbation theory to obtain higher order approximations of this result, using for instance the exact formulas obtained in \cite{Caravellietal}.

\end{document}